\documentclass[reprint,amsmath,amssymb, aps, prl]{revtex4-1}
\usepackage{graphicx}
\usepackage{caption}
\usepackage{subcaption}
\usepackage{epsfig}
\usepackage{color}
\usepackage{amsmath}
\usepackage{amsfonts}
\usepackage{mathrsfs}
\usepackage{txfonts}
\usepackage{color}

\begin{document}
\title{Spinor Condensates on a Cylindrical Surface in Synthetic Gauge Fields} 

\author{Tin-Lun Ho$^{\dag \ast}$ and Biao Huang$^{\dag}$}
\affiliation{$^{\dag}$Department of Physics, The Ohio State University, Columbus, OH 43210, USA\\
$^{\ast}$ Institute for Advanced Study, Tsinghua University, Beijing 100084, China}
\date{\today}

\begin{abstract}
We point out that by modifying the setup of a recent experiment that generates a Dirac String\cite{Hall}, one can create a quasi 2D spinor Bose condensate on a cylindrical surface with a synthetic magnetic field pointing radially outward from the cylindrical surface. The synthetic magnetic field takes the form of the Landua gauge. It is generated by the Berry's phase of a spin texture, frozen by an external quadrupolar magnetic field. Unlike in the planar case, there are two types of vortices (called A and B) with the same vorticity. The ground state for $5\le S\le 9$ consists of a row of alternating AB vortices lying at the equatorial circle of the cylinder. For higher values of $S$, the A and B vortices split into two rows and are displaced from each other along the cylindrical axis $z$. The fact that many properties of a BEC are altered in a cylindrical surface implies many rich phenomena will emerge for ground states in curved surfaces. 
 \end{abstract}

\maketitle

In the study of quantum matter, one usually deals with Euclidean space. Spaces with non-zero curvatures are seldom encountered. Yet  in theoretical studies, torus and spherical surfaces are often used  for computational convenience, or to  
demonstrate hidden topological properties of the system. In recent years, it is found that curvature effects can mimic those of 
gauge fields (as in the case of graphene\cite{graphene}), and can lead to topological responses such as Hall viscosity\cite{hallviscosity}. 
The fact that spatial curvature can help uncover new features of quantum matter makes it desirable to 
create manifolds of controllable curvature, and to develop capability to add in synthetic gauge fields. 

The purpose of this paper is to point out ways to create quantum gases in curved surfaces and to manufacture synthetic gauge fields in such systems. As a first step, we shall consider the simple geometry of an annulus and the case of spinor condensate of bosons with large spin.  We consider Bose-Einstein condensates (BEC) because its quantum properties are often  magnified considerably by Bose statistics.  Large spins particles are considered  because the Berry phase that determines the strength of synthetic gauge fields for these particles is  proportional to their spin.   As we shall see, the system we construct amounts to a system of "charged" bosons in a magnetic field written in Landau gauge. The effective magnetic field pointing radially outward, normal to the cylindrical surface. This gauge field is caused by a spin texture generated by a (real) quadrupolar magnetic field. We shall also point out that the techniques we apply in our construction have all been achieved experimentally; therefore, creating BEC on cylindrical surface with a Landau gauge is feasible. 

As it turns out, for typical parameters of atomic gases, a ground state in the lowest Landau level (``mean field quantum Hall regime'') can only be achieved with  a thousand of particles or less.   A Bose gas with with 10$^{5}$ bosons will form a condensate occupying many Landau levels. For a sufficiently strong synthetic magnetic field, the ground state will contain vortices\cite{fetterrmp}.  However, due to the topological constraint of the annulus and the curvature of the cylinder, the flow pattern of a single vortex as well as the vortex lattice are very different from those found in planar geometry.  For a BEC in a plane, isolated vortices have cylindrical symmetry and the vortex array of a rotating BEC is a hexagonal lattice. 
For BECs in an annulus, however, there are two kinds of vortices with the same vorticity but different flow pattern. In a large effective magnetic field,  the vortices will first align in a single row at the center of the annulus at $z_j=0$ in the azimuthal direction $\varphi$.  After certain proliferation, the vortices split into two lines at $z_j = \pm Z$. The velocity profile resembles that of two counter-circulating superfluid rings separated in the $z$ direction. In the time of flight experiments, 
a row of $n$ vortices will lead to an $n$-fold symmetry in the density profile in the azimuthal direction.  The presence of many quantum phenomena have taken new forms in curved space even for the simple case of Bose condensates suggest a great deal more new phenomena lay in store for more complex quantum systems. 

\begin{figure}[h]
\parbox{4.5cm}{
\includegraphics[width=4cm,height=5cm]{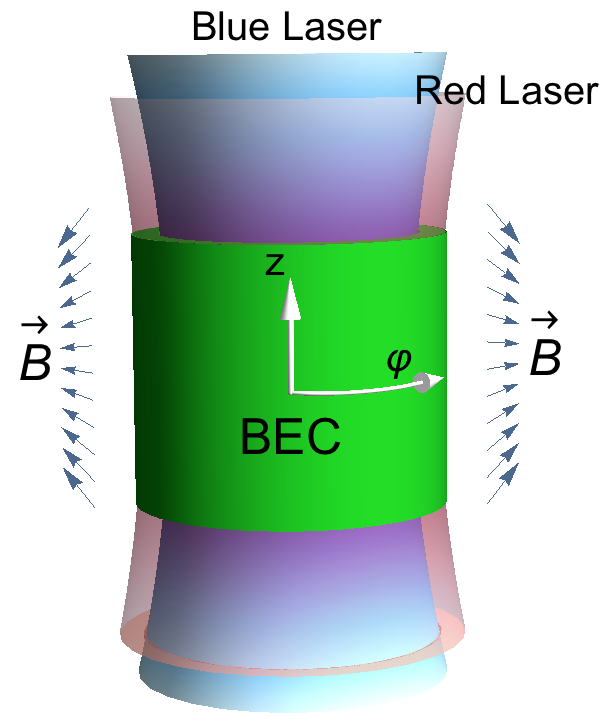}
}
\parbox{4cm}{
\includegraphics[width=4cm,height=3cm]{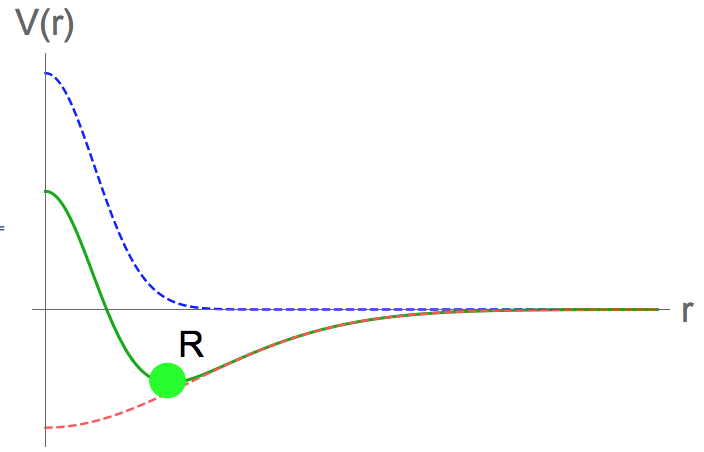}\\
\includegraphics[width=3cm,height=2.5cm]{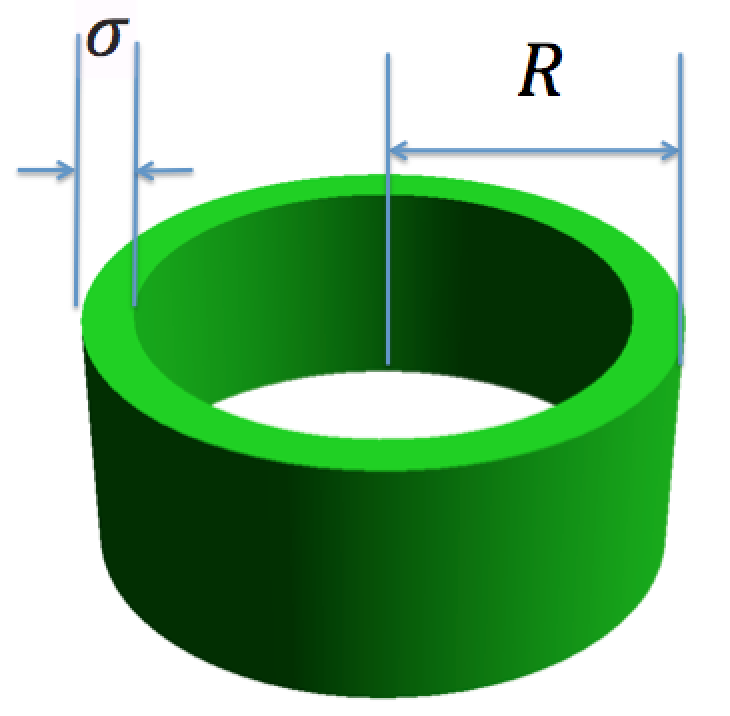}
}
\caption{Schematic plot of the system.\label{scheme}}
\end{figure}

{\em I. Experimental setup and realization of the Landau gauge in a cylinder :} 
Our setup is a modification of the recent experiment by Daivd Hall's group to generate a Dirac string in a BEC by inserting in it a quadrupolar magnetic field\cite{Hall}. Our system is shown in Fig. \ref{scheme}. A confining potential of the form an annulus is constructed by piercing through an attractive potential (produced by a red-detuned laser) with a repulsive core (produced by a blue-detuned laser). This will create a confining well in the radial direction with a minimum at radius $R$. A  harmonic potential $V(z) = M\omega_z^2 z^2/2$  is imposed along the $z$-direction, which  is far weaker than the harmonic trap in the radial direction in the neighborhood of $R$. When a quantum gas is loaded onto this trap, it will form a cylindrical layer with radius $R$ and thickness $\sigma$. By increasing $R$ and reducing the frequency $\omega_{z}$, 
we shall have $R\gg \sigma$. The quantum gas then becomes a curved quasi-2D system.

Next, we insert a quadrupole magnetic ${\bf B}$ at the center of the cylinder, 
\begin{eqnarray}\nonumber
\mathbf{B}(\mathbf{x}) &=& B_0 (x\hat{x}+y\hat{y}-2z\hat{z}) \\ \label{bdirection}
&=& B_0 r(\sin\theta\cos\varphi \hat{x} +\sin\theta\sin\varphi \hat{y}-2\cos\theta\hat{z}).
\end{eqnarray}
The direction of ${\bf B}$  on the cylindrical surface is shown in Fig. \ref{scheme}. This quadrupole field is introduced to generate an effective gauge field for the bosons that are spin aligned with the local field. 
The insertion of a quadrupolar magnetic field into a 3D BEC  has been recently performed in the experiment in Ref.\cite{Hall}. 
Our configuration is basically a modification of the set up in Ref.\cite{Hall} there by piercing through their harmonic potential with a repulsive blue-detuned laser. 
%Recently, this idea was successfully executed  by several experimental groups to realize the synthetic monople field, using the external quadrupolar field. {\color{red} [Choi (2012 skyrmion),(2013)] \& [David Hall(2014)]}.  {\color{red} [Hall]} directly observed the Dirac string \cite{dirac} and the monopole center in a  three dimensional condensate; while {\color{red} [Choi]} confined the system in two dimensions, and obtained skyrmion textures. In {\color{red} [Choi]}, it was realized that the spin winding within the finite planar condensate, which corresponds to synthetic flux,  was not enough to support vortex ground states. So additonal shaking of the quadrupole center  was applied to impart angular momentum to the system. 
 For a sufficiently large quadrupole field, the spins of the bosons will align with the local magnetic field. Denoting direction of the spin as $\hat{l} = \cos\alpha_1 \hat{z} + \sin\alpha(\cos\beta\hat{x} + \sin\beta \hat{y})$, Eq.(\ref{bdirection}) implies 
 \begin{equation}
 \alpha = \varphi , \qquad
 \cos\beta = -\frac{2z}{\sqrt{R^2+4z^2}}.
 \end{equation}
The condensate wave function of bosons with spin $S$ is then  $\psi_{a}(\mathbf{x}) = \zeta_{a}(\mathbf{x})\phi(\mathbf{x})$, where $a$ is the spin index,  
$\zeta_{a}(\mathbf{x})$ is a normalized vector aligned with the local magnetic field, i.e. $\hat{\bf B}(\mathbf{x})\cdot {\bf S}_{ab} \zeta_{b}(\mathbf{x})= S\zeta_{a}(\mathbf{x})$. With the spin direction frozen by ${\bf B}$,  
the kinetic energy becomes 
\begin{equation}
|\nabla\psi|^2 =\left |\left (\frac{\nabla}{i}+\frac{\zeta^\dagger\nabla\zeta}{i}\right)\phi \right|^2 + |\nabla\zeta|^2 |\phi|^2 + (\zeta^\dagger\nabla\zeta)^2 |\phi|^2. 
\label{KE} \end{equation}
The field $\phi$ experiences an effective gauge field $\mathbf{A}(\mathbf{x}) \equiv -i \zeta^\dagger(\mathbf{x})\nabla \zeta (\mathbf{x})$. 
The energy functional of $\psi$ (with spin indices suppressed), 
\begin{equation}\label{energyfunc}
E[\psi] = \int d^2x  \left[ \frac{\hbar^2}{2M}|\nabla \psi|^2- \left(\mu - V(z)\right) |\psi|^2+ \frac{g}{2}|\psi|^4 \right],
\end{equation}
reduces to a functional of $\phi$, and the last two terms in Eq.(\ref{KE}) (which we shall see is only a function of $z$) can be absorbed into potential $V(z)$. 
%where the summation over spin indices $\mu$ is understood. 
%Plug (\ref{psidecompose}) into (\ref{energyfunc}), we have for the kinetic term \cite{deriveh}
%\begin{equation} |\nabla\psi|^2 =\left |\left (\frac{\nabla}{i}+\frac{\zeta^\dagger\nabla\zeta}{i}\right)\phi \right|^2 + |\nabla\zeta|^2 |\phi|^2 + (\zeta^\dagger\nabla\zeta)^2 |\phi|^2. \end{equation}
%Near $z=0$, the last two terms can be obsorbed into the harmonic trap potential \cite{deriveh}, and therefore can be omitted. 
In the limit where the radius $R$ of the quantum gas is much larger than its thickness $\sigma$, the system can be regarded as a quasi-2D system, and the field $\phi$ is a function of $z$ and the azimuthal angle $\varphi$. It is then straightforward to show that 
%From the first term we see that the gauge potential  $\mathbf{A}(\mathbf{x}) \equiv -i \zeta^\dagger(\mathbf{x})\nabla \zeta (\mathbf{x})$  comes from the spin texture. In our setting \cite{derivespin},
\begin{eqnarray}
\nabla\phi &=& \hat{z}\partial_z \phi + \frac{\hat{\varphi}}{R} \partial_\varphi \phi,\\
-i\zeta^\dagger\nabla\zeta &=&  -S(\nabla\alpha) \cos\beta = \frac{\hat{\varphi}}{R}\frac{2zS}{\sqrt{R^2 + 4z^2}} \approx \frac{\hat{\varphi}}{R} \frac{2zS}{R}   \label{A}\\
\nabla \times {\bf A} &=&  \left( R^{-1}\partial_{\varphi} A_{z} - \partial_{z} A_{\varphi}\right) \hat{\bf r} = - (2S/R^2)\hat{\bf r}, \qquad
\end{eqnarray}
for $z<R/2$, 
where $\hat{\bf r}$ is the radial unit vector in cylindrical coordinate. 
This effective magnetic field corresponds to $2S$ unit of phase winding in the planar case. Thus, for Bose gases with sufficiently large spin, the ground state is expected to contain vortices. 

To simplify notations, we measure length in units of $R$, so that  $z/R \rightarrow z$, and $z$ is now dimensionless. 
% R\varphi \rightarrow \varphi$. 
 We further introduce the dimensionless variables $\tilde{\alpha} \equiv M\bar{\omega_z}R^2/\hbar, \tilde{\mu}\equiv \mu/(\hbar^2/2MR^2), \tilde{g}\equiv g/(\hbar^2/2MR^2)$, 
then the energy functional becomes
\begin{eqnarray}\nonumber
&&E[\psi] = \frac{\hbar^2}{2MR^2}
\int d^2 x\,\, {\cal E},
\\ \label{etot}
{\cal E} &=& 
\left|\partial_z\phi\right|^2 + \left|\left(-i\partial_\varphi + 2Sz\right)\phi\right|^2
- \left(\tilde\mu - \tilde{\alpha}^2z^2 \right) |\phi|^2  +  \frac{\tilde g}{2}|\phi|^4.\quad
\end{eqnarray}
The first two terms represent a system in uniform magnetic field with a Landau gauge. 

{\em II. Isolated vortex and vortex array patterns on a cylinder:}
To find the vortex pattern for the cylindrical 2D condensate, we seek for a variation solution in the Thomas-Fermi (TF) limit for the functional Eq.(\ref{etot}). We use the variational anzatz
% seek for vortex solution to this problem in the Thomas-Fermi (TF) limit, using the variational wavefunction
\begin{equation}
\phi = \sqrt{n} f e^{i\Theta}
\label{vari} \end{equation}
Here $n(\varphi, z)$ is the density profile given by the TF approximation in the absence of vortices. $f(\varphi,z)$  punches hole in the vortex locations $(\varphi_i, z_i)$ and modifies the density profile,
\begin{equation}
f(\varphi, z) = \prod_i \tanh \frac{\sqrt{(\varphi-\varphi_i)^2+(z-z_i)^2}}{\xi},
\label{f} \end{equation}
 where $\xi$ is the core size, also written in units of $R$. The phase $\Theta(\varphi, z)$ gives 
%\begin{equation} h\equiv \exp(i\Theta) \end{equation} 
a $2\pi$ phase winding around each vortex. This form has been shown to match well with experiment in the case of rotating gases\cite{fetterrmp}. 

\begin{figure}[h]
\begin{center}
\begin{subfigure}{3cm}
\includegraphics[width=2cm]{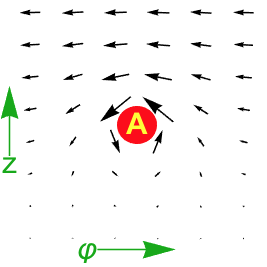}
\caption{A-vortex.} \label{fig:va}
\end{subfigure}
\begin{subfigure}{3cm}
\includegraphics[width=2cm]{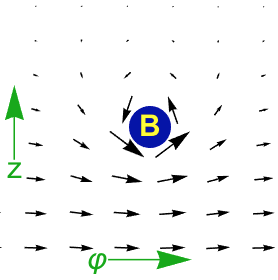}\\
\caption{B-vortex}\label{fig:vb}
\end{subfigure}
\\
\begin{subfigure}{9cm}
\includegraphics[width=8.5cm]{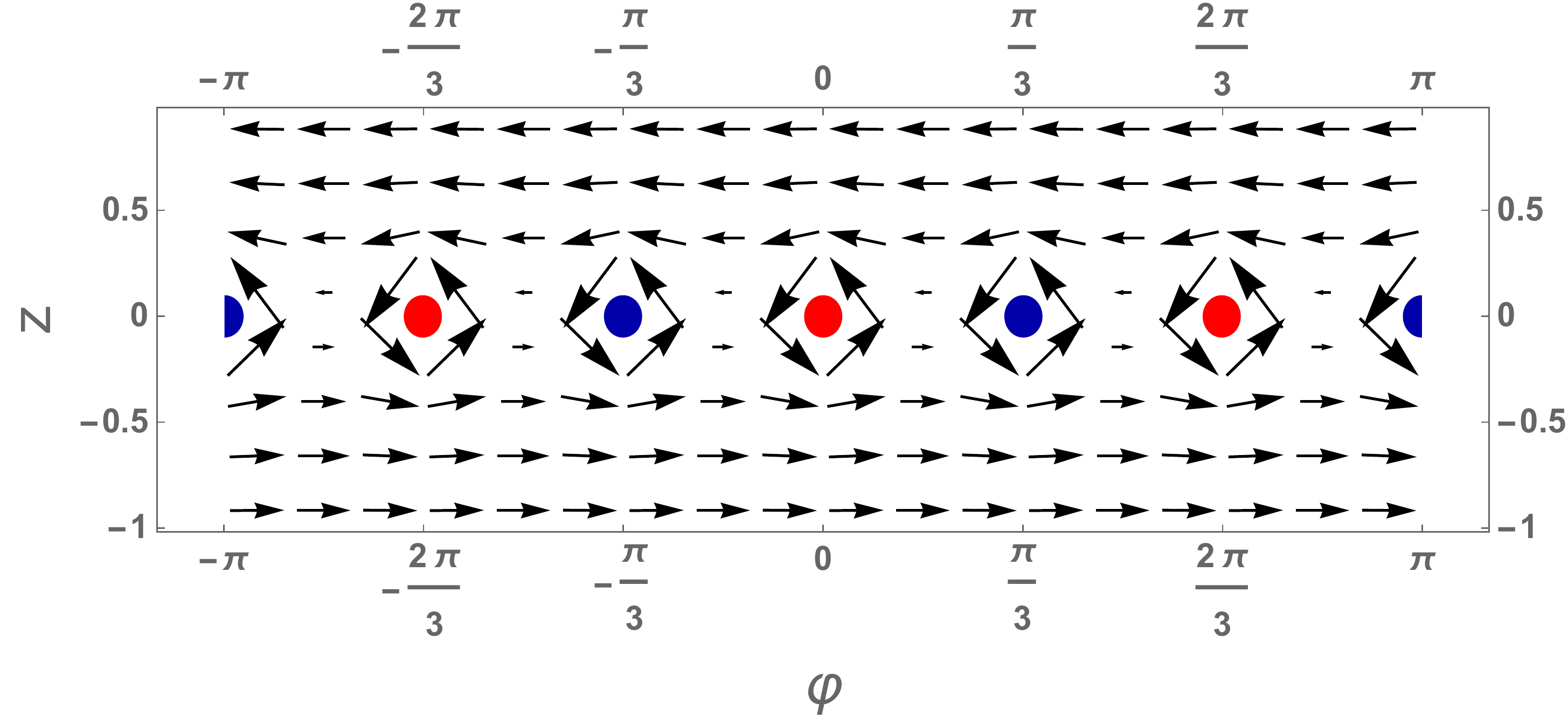}\\
\caption{6 alternating vortices aligned along $z=0$.} \label{fig:6v}
\end{subfigure}
\end{center}
\caption{Velocity profiles for two types of vortices and the vortex array.}
\end{figure}

The periodicity of the wave function (or $ e^{i\Theta}$)
 in $\varphi$, however, strongly constraints the form of $\Theta$. From the conformal mapping between a punctured plane and a cylinder, one finds two distinct phase factors $e^{i\Theta^{\pm}_{j}}$ that have the same $+2\pi$ phase winding about the local 
 $(\varphi_{j}, z_{j})$,  
 \begin{equation}
%\Theta^\pm(\varphi, z) = \sum_j \Theta^\pm_j(\varphi, z),\qquad
{\rm exp}(i\Theta^\pm_j) = \frac{W^{\pm}}{|W^{\pm}|},   \,\,\,\,\,\,\,\,\, W^{\pm}= e^{\pm iu}-e^{\pm iu_j}  
%\frac{e^{\mp iu}-e^{\mp iu_j}}{|e^{\mp iu}-e^{\mp iu_j}|},
\end{equation}
where $u=\varphi + iz$, and $ u_j = \varphi_j  + iz_j$ is the vortex location.
The superfluid velocity of each vortex is  $ \frac{\hbar}{m}\nabla\Theta^\pm$
 %$\mathbf{v}_s^{A,B} = \frac{\hbar}{m}\nabla\Theta^\pm$, 
 where
\begin{eqnarray}
\partial_\varphi\Theta_j^\pm &=&\mp  \frac{e^{\pm(z-z_j)}-\cos(\varphi-\varphi_j)}{ 2\left[\cosh(z-z_j)-\cos(\varphi-\varphi_j)\right]},\\
\partial_z\Theta_j^\pm &=& \frac{\sin(\varphi-\varphi_j)}{  2\left[\cosh(z-z_j)-\cos(\varphi-\varphi_j)\right]}.
\end{eqnarray}
One can easily identify that $\nabla\Theta^+_j$ is related to $\nabla\Theta^-_j$ by a $\pi$ rotation with respect to the vortex core. 
For $|z-z_j|\gg0$, they approach
\begin{equation}
\nabla\Theta^+_j \rightarrow
\left\{
\begin{array}{rl}
-1\cdot \hat\varphi, & z\gg z_j\\
0, & z\ll z_j
\end{array}
\right. ,
\quad
\nabla\Theta^-_j \rightarrow
\left\{
\begin{array}{rl}
0 , & z\gg z_j\\
1\cdot\hat\varphi, & z\ll z_j \,\,\,\, .
\end{array}
\right.
\end{equation}
So, as a function of $z$, $\partial_\varphi \Theta_j^{+}$ (or $\partial_\varphi \Theta_j^{-}$) approaches to a constant far above (or far below) the vortex core, and vanishes on the other side, as shown in Fig. \ref{fig:va} and Fig. \ref{fig:vb}. Because of this feature, we call $\Theta_j^+$  and $\Theta_j^-$ 
the A- and B-vortex as their superfluid velocities are mostly non-vanishing "above" and "below" the vortex core respectively. 
Note that despite their very different velocity patterns, both of them have the same $+2\pi$ circulation around the vortex core. 
In the case  we have  A-vortices on a set of $Q$ point $\{ u_{j}, j=1,..Q \}$ and B-vortices on another set of $Q$ points, $\{ u_{j'}, j=Q+1,..2Q \}$, the phase factor is 
\begin{equation}
{\rm exp}(i \Theta)= W/|W|, \,\,\,\,\,\,\,W = \prod_{i=1}^{Q} W^{+}_{j}  \prod_{j=Q+1}^{2Q} W^{-}_{j'} .
\label{W} \end{equation}

\begin{figure}[h]
\begin{center}
\includegraphics[width=7.5cm]{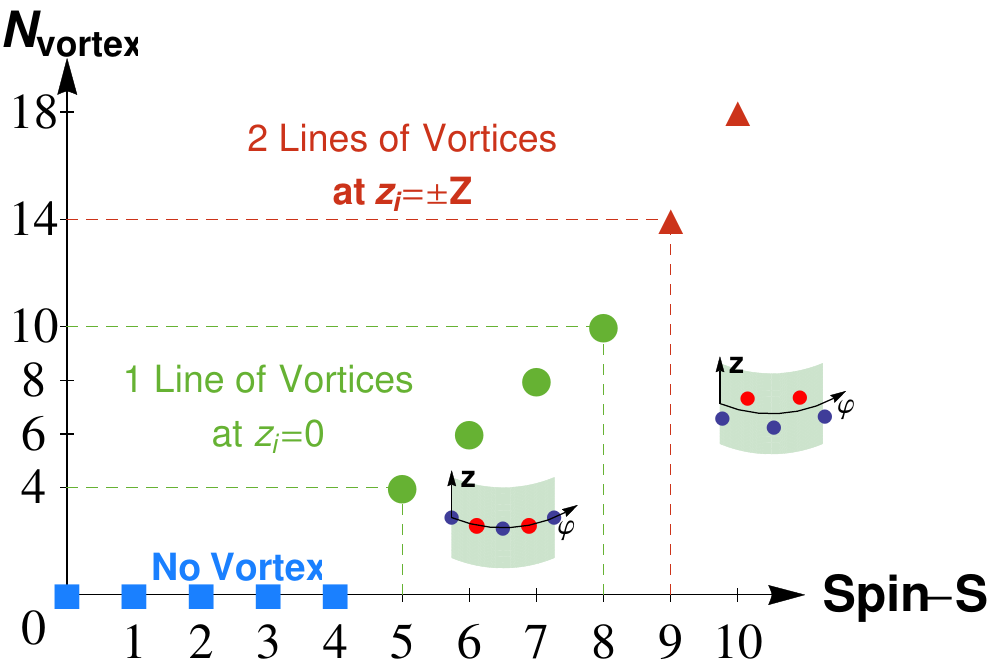}\\
\caption{Number of vortices versus spin in the ground state. For $S\le 4$, the ground state contains no vortex. Within $S=5,6,7,8$, the vortices aligned in one row at $z_i = 0$ with the pattern A-B-A-B. For $S\ge 9$, the vortices array splits into two rows centering at $z_i = \pm Z$, with all A vortices aligned in one row and all B vortices in another.}\label{fig:scale}
\end{center}
\end{figure}

We have minimized the energy Eq.(\ref{etot}) with the variational wave function Eq.(\ref{vari}), where the vortices at $(\varphi_j, z_j)$ can be either A or B type. We shall first present our results and describe the details of our calculation at the end. 
% with more details given in supplementary materials. 
 The variational result is shown in Fig. \ref{fig:scale}. First, since the system has reflection symmetry in z-direction, A and B vortices must appear in pairs at appropriate location to respect this symmetry.  Our variational calculation shows that for $S\le 4$, the strength of the gauge field is not strong enough to generate vortices in the ground state.  For $S=5,6,7,8$, there are $4, 6, 8, 10$ vortices respectively lying on   the circle at  $z=0$  (i.e. $z_j=0$).  These rows of vortices are all in the alternating pattern A-B-A-B-.. . 
 For $S\geq 9$, the vortices split into two rows above and blow $z=0$, with the A-vortices shifted up and B-vortices shifted down. 
 
 In our calculations, we have considered a system of $10^5$ bosons, with a chemical potential  such that the extent of the condensate along $z$ is of the order but shorter than $R$ to stay so that ${\bf A}$ has the form of the Landau gauge, 
 Eq.(\ref{A}). The size of the vortex core is chosen to be the coherence length of uniform condensate with the chemical potential at $z=0$.  
%In the case of $2Q$ vortices, i.e. $i=1$ to $2Q$ in Eq.(\ref{f}), the way that we attach A-vortices to a set of points $\{ u_{j}\}$ $j=1$ to $Q$, and B-vortices to another set $\{ u_{j'}\}$, $j'=Q+1$ to $2Q$ is to multiply the function $\sqrt{n}f$ in Eq.(\ref{vary}) by the factor ${\rm exp}(i\Theta) = W/|W|$, where 
% \ The way that we attach A-vortices to a set of points $\{ u_{j}\}$ $j=1$ to $Q$, and B-vortices to another set of points $\{ u_{j'}\}$, $j'=Q+1$ to $2Q$  to multiply the function $\sqrt{n}f$ in  With this variational form, most calculations of the energy functional Eq.(\ref{etot}) can be handled analytically, and the energy becomes a function of the locations of the vortices once the choice of the vortex type (i.e. A or B) is made. The detail expressions of the energy is given in the supplementary materials. 
 
 We conclude this section by examining the phase function and velocity field of the alternating vortex row 
 in greater detail.  Let us consider the case of $S=6$ (corresponding to $^{168}$Er) where the ground state has 6 vortices of alternating A and B type equally spaced on the circle at $z=0$, say, at $\varphi =  n \frac{2\pi}{6}$, where $n= 0,1,..5$.
Defining $w=e^{iu}=e^{i\varphi -z}$, and $\alpha= e^{2\pi/6}$, 
the phase function $W$ in Eq.(\ref{W}) is then
% $\varphi = 0, \pi, \pm \frac{\pi}{3}, \pm \frac{2\pi}{3}$.
%  $z=0$ at $\varphi = 0, \frac{2\pi}{6},  \frac{4\pi}{6}, \frac{6\pi}{6}, \frac{8\pi}{6}, \frac{10\pi}{6}$.
%\begin{eqnarray} W&=& (w-1)(w^{-1}-\alpha^{-1}) (w- \alpha^2)(w^{-1}-\alpha^{-3})(w-\alpha^4)(w^{-1}-\alpha^{-5})\\ &=&w^3-w^{-3}
 %\end{eqnarray}
 \begin{eqnarray}
 W&=& \prod_{n=0,1,2} \left[ (w-\alpha^{2n})(w^{-1}-\alpha^{-(2n+1)}) \right] \\
 &=&w^3-w^{-3}
 \end{eqnarray}
The wave function is simply 
\begin{equation}
\phi(\varphi, z )\sim  \sqrt{1-z^2}   f(\varphi,z) \frac{e^{3i\varphi -3z}-e^{-3i\varphi+3z}}{|e^{3i\varphi-3z}-e^{-3i\varphi+3z}|},
\label{6v} \end{equation}
Eq.(\ref{6v}) shows that for $z>0$ ($z<0$),  $\phi(\varphi, z )$ quickly approaches $e^{-3i\varphi}$ ( $e^{3i\varphi}$).  The system then consists of two counter circulating superflow above and below $z=0$, as shown in Fig.2(c). 

The reason that the vortex pattern is so different from the planar case is a consequence of the confined geometry. As one increases the strength of the gauge field by, say, increasing $S$, more vortices will come. At some point, it will be too costly to put all the vortices in a single row.  Our calculation shows that the vortices will separate into two rows, as shown in Fig.3 for the cases $S\geq 9$. Mathematically, if one squeeze in a large number of vortices by choosing a very large value of $S$, the vortices will then organize in a lattice on a cylinder as in the planar case.  The separation of vortices in two rows as shown in Fig.3 can be viewed as a tendency towards the lattice limit.

\begin{figure}[h]
\begin{center}
\includegraphics[width=6cm]{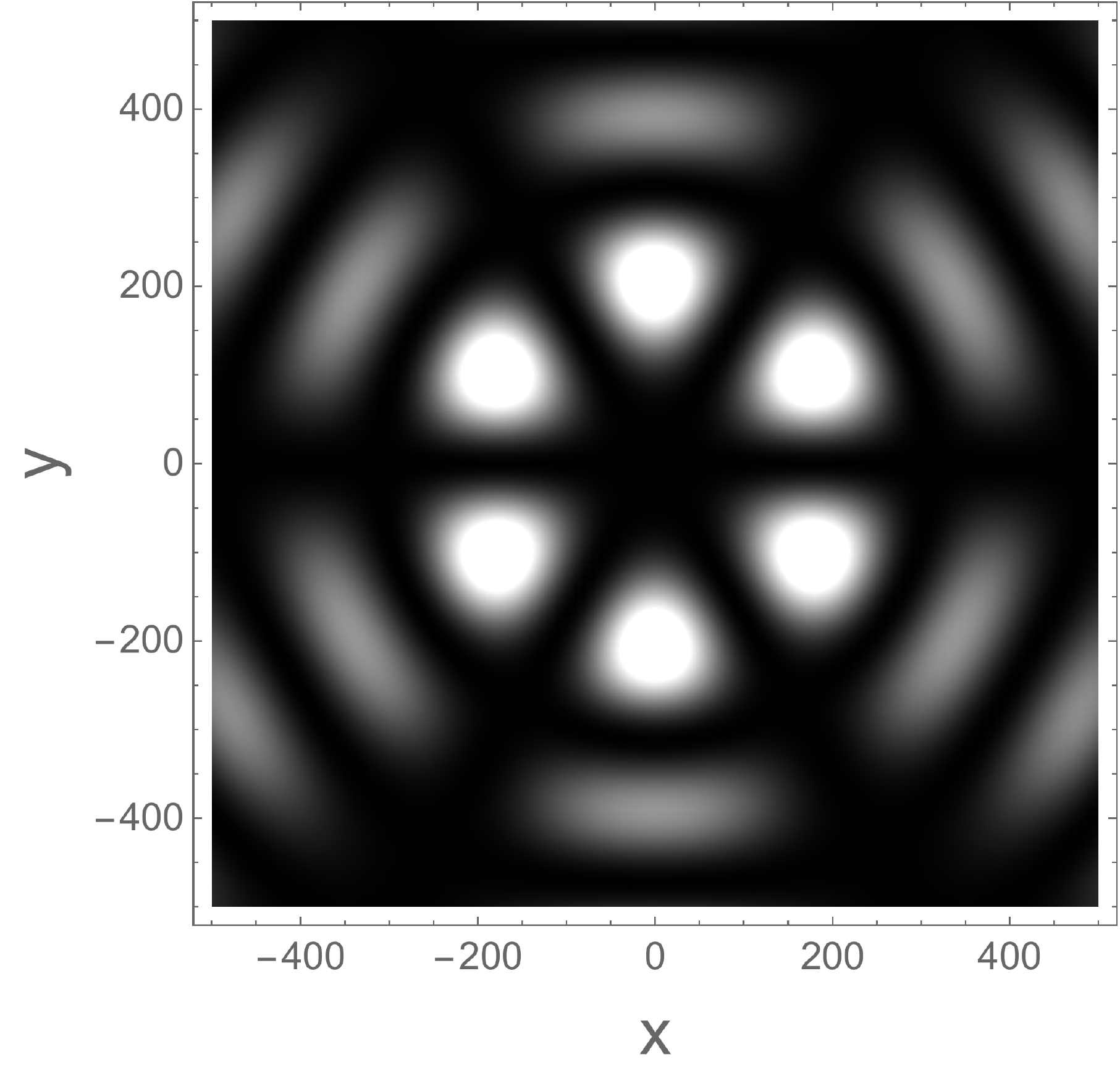}
\end{center}
\caption{Density profile at $x_0/R \equiv \sqrt{\hbar t/m}/R = 7$ on the $x$-$y$ plane for $z=0$. Length is measured in units of $R$.}\label{fig:6v}
\end{figure}

{\em III. Signature of the vortex array:} The presence of these alternating vortex array can easily be detected in time of flight experiments. As we have discussed, the AB vortex row consist of $2n$ vortices will generate two counter phase superflow  
$e^{-in\varphi}$ and $e^{+in\varphi}$ 
for $z>0$ and $z<0$. The system can be approximated by two rings of condensates with opposite circulation, with one ring sitting above the other along $z$ with a separation or the order of their radius $R$.  
In the time of flight experiment, these two rings will produce an interference at the $z=0$ plane of the form of  
$e^{-in\varphi}+ e^{+in\varphi}$ and therefore exhibit a density pattern with $2n$-fold symmetry. 
This effect is in fact found in an explicit calculation of  the ballistic expansion of the vortex row condensate in Eq. (\ref{6v}). The time evolution of the condensate is given by 
 \begin{equation}
 \phi(\mathbf{x},t) = \int_{-1}^1 dz \int_{-\pi}^\pi d\varphi U(\varphi,z,t; \varphi',z') \phi(\varphi',z').
 \end{equation}
 where $U(\varphi, z, t; \varphi',z')$ is the Green's function for free particle propagation at large distance and at long times in cylindrical coordinates,  $U(\varphi, z, t; \varphi',z') \approx 
 \exp\left[ -i( Rr\cos(\varphi-\varphi') + zz')/x_0^2
 \right]$, and  $x_0 = \sqrt{\hbar t/m}$.  
 The density pattern at the equatorial plane $z=0$ at long times is shown in Fig.4.

 {\em IV Details of the variational calculation:}
 Since both $n, f$ in Eq.(\ref{vari}) are real and $|h|=1$, we can rewrite Eq.(\ref{etot}) as 
\begin{eqnarray}
{\cal E} &=& {\cal E}_1 + {\cal E}_2 + {\cal E}_3,\\ \label{e1}
{\cal E}_1 &=& \left[ (\partial_z\Theta)^2 + (\partial_\varphi\Theta + 2Sz)^2 \right] nf^2\\ \label{e2}
{\cal E}_2 &=& -nf\nabla^2 f - \tilde{g} n^2 (f^2- \frac{f^4}{2}-\frac{1}{2})- \frac{\tilde{g}n^2}{2}\\ \label{e3}
{\cal E}_3 &=& \left[ (\nabla \sqrt{n})^2 - (\mu-V(z))n + \frac{\tilde{g}}{2}n^2\right] f^2
\end{eqnarray}

We  recognize ${\cal E}_3$ has the form of usual Gross-Pitaevskii energy functional, and determines the density profile in the absence of vortices. 
Applying TF approximation to Equation (\ref{e3}) and vary with respect to $n$, we obtain
\begin{equation}
n(z) = \frac{\mu-V(z)}{g} = \frac{M\omega^2 R^2}{2g}(\beta^2-z^2),
\end{equation}
with the constraint $n\ge0$. Here  $\beta = \sqrt{2\mu/M\omega_z^2 R^2}$, and the condensate spreads within $(-\beta, \beta)$. Taking the condensate to be homogeneous along radial direction, the total number of particles given by this density profile is then 
\begin{equation}
N = \frac{\beta^3}{3}\times \frac{\tilde{\alpha}^2}{a_s/\sigma},
\end{equation}
 where $a_s$ is the scattering length related to $g$ by $g=\frac{4\pi\hbar^2 a_s}{M}$. 

Further, in the absence of vortices, $f=1$ everywhere, and the first two terms in ${\cal E}_2$ vanishes. That means ${\cal E}_2$ is the energy increase due to the existence of vortices. To analyze it, we first notice that the core size is given by the coherent length $\xi \sim \frac{\hbar}{\sqrt{2Mn(z) g}}\frac{1}{R}$. Taking the value at $z=0$, we have $\xi_0 \sim \frac{1}{\tilde{\alpha}}$. Note $\tilde{\alpha} = R^2/d^2$, where $d = \frac{\hbar}{M\omega_z}$ is the natural length of the harmonic trap along $z$, which  we take to be an order of magnitude smaller than $R$. Then $\xi\sim 10^{-2}\ll 1$. That means the first two terms in  ${\cal E}_2$ is only non-zero close to each vortex core. So we can make approximations using the identity
$\int_{a_1}^{a_2} dx f_1(x)f_2(x) = (\sum_i f_1(x_i)) \int_{-\infty}^\infty dx f_2(x)$, where $f_1(x)$ is a smooth function, and $f_2(x)$ is only non-zero within a small region around $x_i$'s.
Then the integration can be done analytically: $
\int d^2 x\,\, {\cal E}_2 = -\frac{2\pi\tilde{\alpha}^2}{3}+ (2+\tilde{\alpha}^2 \xi^2) C\sum_i (\beta^2-z_i^2)$,
where $C=\frac{\pi}{6}(4\ln2 -1)\approx 0.928125$ is the integration result, and $z_i$ is the $z$ coordinates of vortices, and it is confined to $|z_i|<\beta$.  As expected, the vortex core energy would like to repel vortices out of the condensate by increasing $|z_i|$. The constant term plays no role and will be omitted. Also, we take $\tilde{\alpha}\xi = 1$ according to the above analysis.

In summary, measured in units of $\frac{M\omega^2R^4}{2g}$, the energy functional is reduced to
\begin{eqnarray}\nonumber
E[\psi] &=& 3C\sum_i(1-z_i^2) 
+
 \int_{-\pi}^\pi d\varphi \int_{-\beta}^\beta dz \\ \label{efinal}
 && \qquad \times \left[ 
(\partial_z\Theta)^2 + (\partial_\varphi\Theta + 2Sz)^2 
\right]  (\beta^2-z^2)f^2.\quad
\end{eqnarray}
For fixed total particle number $N$, reducing $\beta$ corresponds to increasing the strength of the trap or weakening the interaction.
We will take $\beta=1$ in the following.
The vortices should induce the superfluid velocity profile that serves to cancel $(2Sz)$ with $\partial_\varphi\Theta$ in the above energy functional, and therefore lower the energy.

{\em Concluding Remarks:} The emergence of two kinds of vortices with identical vorticity in a cylindrical manifold is a new feature of Bose condensates in a cylinder. It is a consequence of the topological constraint on the single value-ness of the wave function (i.e. that forces the spatial dependence to be expressed in terms of $e^{in\varphi}$) and should persists even when the manifold is deformed. Although we focus on a particular aspect of the quantum gas in curved surfaces, there are a lot more to explore especially 
for systems with more complexity.  Realization of quantum gases in curved surfaces will surely open an exciting direction for cold atom research.

{\em Acknowledgments:} This work is supported by the NSF grant DMR-1309615, the 
MURI grant FP054294-D, and the NASA grant 1501430.

\end{document}